\documentclass[a4paper]{article}
\usepackage{amsfonts}
\usepackage{amssymb}
\usepackage{amsthm}
\usepackage[margin=2cm]{geometry}
\usepackage[dvips]{graphicx}
\usepackage{pgf}
\usepackage[T1]{fontenc}
\usepackage[latin2]{inputenc}
\usepackage{subfigure}
\usepackage{amsmath}
\usepackage{bbm}

\begin{document}

%\draft
%\twocolumn
\title{On von Neumann and Bell theorems applied to quantumness tests \\
%Response to Marek \.Zukowski paper , arXiv:0809.0115v1
}

\author{Robert Alicki \\ 
  {\small
Institute of Theoretical Physics and Astrophysics, University
of Gda\'nsk,  Wita Stwosza 57, PL 80-952 Gda\'nsk, Poland}\\
}

\date{\today}
% \date{July 26, 2003}
\maketitle

\begin{abstract}
The issues, raised in arXiv:0809.011, concerning the relevance of  the von Neumann  theorem for the single-system's quantumness test proposed in arXiv:0704.1962 and performed for the case of single photon polarization
in arXiv:0804.1646, and the usefulness of Bell's inequality for testing the idea of macroscopic quantum systems  are discussed in some details. 
Finally, the proper quantum mechanical description of the experiment with polarized photon beams
is presented. 
\end{abstract}
\section{Introduction}
The recent paper of Marek \.Zukowski \cite{MZ} is aimed as a criticism of the paper by myself and Van Ryn \cite{AV} devoted to " a simple test of quantumness  for a single  system" which was  experimentally realized in \cite{B}. The ideas of \cite{AV} were further developed in \cite{APV} which appeared after \cite{MZ} has been finished.
\par
The three main thesis of \cite{MZ} are the following:\\
I)  "... simple test of quantumness for a single system....has exactly the same relation to the discussion...as the von Neumann theorem"\\
II) "As far as a \emph{direct} detection of non existence of \emph{any} classical probabilistic models is concerned we are left with the two theorems of Bell..."\\
III) "..., the example given in \cite{AV} and realized in \cite{B} does have a classical model, like every second order (in terms of fields) photonic interference effect. The observed phenomena can be interpreted as non-classical only due to the statistical properties of the parametric down conversion process".
\par
Although this kind of problems attracted attention for several decades and stimulated many authors to produce enormous number of publications it seems that the serious confusions and misconceptions are still alive and need further debate and clarification (see an excellent discussion in \cite{Str}). Recently, the problem of "quantumness" became important for more practical reason, namely, the question of (non)existence of \emph{macroscopic quantum systems} which could be used as implementations of qubits in quantum information processing.

\section{Two theorems and related quantumness tests}
To eliminate the main source of confusion  in the debate one should make the distinction between two different classes of problems:\\
Q1) Impossibility of description of a given set of experimental data by a classical probabilistic model.\\
Q2) Impossibility of a local realistic interpretation of quantum mechanics.\\
The papers \cite{AV, APV} concern the problem Q1 while the arguments of \cite{MZ} are based on the rich family of results known under collective name of Bell's Theorem \cite{stanford} and concerning Q2.  Firstly, Q1 and Q2 are independent from the logical point of view. Secondly,  in Q1 the classical probabilistic model means a theory like statistical classical mechanics with observables forming an algebra of functions and states being all probability distributions on a certain "phase space". On the other hand, in Q2 the notion of  local realistic model is often formulated in philosophical language leaving a space for different interpretations in precise mathematical terms \cite{phil, Str}. Nevertheless, both Q1 and Q2 have something in common. This is the search for the particular features  of  classical probabilistic models which are not present in the quantum formalism.
For the purpose of further discussions I present two theorems formulated in the algebraic language and concerning this type of features. In the following ${\cal A}$ denotes a $C^*$-algebra with the identity $I$ and provides a model for all  bounded observables of the physical systems as self-adjoint elements of ${\cal A}$. The set of all linear, positive and normalized functionals on ${\cal A}$ denoted by ${\cal S}({\cal A})$ is identified with all physical states of the system. The mean value of the observable $A$ in the state $\rho$ is given by $\rho(A)$. The spectrum $\mathrm{Sp}(A)$ consists of all numbers $a$ (real for self-adjoint $A$) for which the element $A-aI$ possesses no inverse in the  algebra ${\cal A}$. One interprets the spectrum $\mathrm{Sp}(A)$ as possible outcomes of the measurement for the observable $A$ \cite{alg}. 

{\bf Theorem A}
For any self-adjoint elements $A,B,C\in {\cal A}$ the following implication \\
\begin{equation}
 \rho (A)+ \rho(B)= \rho(C)\quad  \mathrm{for\ all} \quad\rho\in{\cal S}({\cal A})\Rightarrow \mathrm{Sp}(C)\subset\mathrm{Sp}(A) + \mathrm{Sp}(B)
\label{thB}
\end{equation}
holds if and only if ${\cal A}$ is commutative and hence isomorphic to an algebra of continuous functions on a certain
compact set.

{\bf Theorem B}
For any self-adjoint elements $A,B\in {\cal A}$ the following implication \\
\begin{equation}
0\leq \rho (A)\leq \rho(B) \mathrm \quad \mathrm{for\ all}\quad \rho\in{\cal S}({\cal A})\Rightarrow \rho (A^2)\leq \rho(B^2)
\label{thA}
\end{equation}
holds if and only if ${\cal A}$ is commutative and hence isomorphic to an algebra of continuous functions on a certain
compact set.

Both theorems follow from the known results in the theory of operator algebras, the equivalence of the assumptions (\ref{thA}) and (\ref{thB}) is also discussed in \cite{MZ}. They can be used as a motivation for two \emph{quantumness tests} which can be applied to a collection of experimental data \cite{obs}.
\par
{\bf QTest A} Find three observables $A, B, C$ which averaged values satisfy the equality 
\begin{equation}
 \rho (A)+ \rho(B)= \rho(C)\quad  \mathrm{for\ all\ states} \quad\rho
\label{tA}
\end{equation}
but the possible outcomes of $C$ are not given by the sums of the outcomes of $A$ and $B$.
\par
{\bf QTest B} Find two observables $A, B$ which averaged values satisfy the inequality 
\begin{equation}
0\leq \rho (A)\leq \rho(B) \quad  \mathrm{for\ all\ states} \quad\rho
\label{tB}
\end{equation}
but for a certain state $\sigma$ the second moments fulfill $\sigma(A^2) > \sigma(B^2)$.

Although no experiment can be conclusive in the philosophical sense (even the existence of the World cannot be proved \cite{Ing}) one can discuss (always a bit subjectively) the level of confidence for the QTestA and  QTestB. In both cases the difficulty is hidden in the words "for all states" as one can never perform experiments for all possible states. If the set of experimentally accessible states is  restricted  too much, it can happen that for the QTestA there exists another observable $C'$ which yields the same mean values as $C$ for the all accessible states but nevertheless possesses as outcomes the sums of outcomes for $A$ and $B$. Similarly, for the QTestB there may exist non accessible states which violate the inequality (\ref{tB}). To remove the later possibility an additional assumption - \emph{minimality of the model} - was introduced in \cite{AV, APV}:\\
\emph{ If for any pair of experimentally accessible observables $A, B$  the inequality (\ref{tB}) is confirmed for all experimentally accessible states, then the same inequality holds for all states in the model.}\\
Now, the positive result of the QTestB excludes the minimal classical model for the experimental data.

The practical and common sense justification of the minimality assumption was already proposed in \cite{AV} and discussed in some details using toy models in \cite{APV}. It is based on a general observation that there is a certain symmetry between state preparation and measurement. Measuring apparatus involves some  selection (filtering) procedures which are used
for state preparation as well. Therefore, for any fixed technological implementation we can assume that the resolution on the side of state preparation is similar to the resolution on the side of measurement \cite{dual}. On the other hand the existence of accessible classical observables $A, B$ for which $B-A$ possesses negative outcomes which are  always averages out by all accessible states means that the resolution of the prepared states is much lower than the resolution of the measurable observables. The similar argument applies to the test A because for classical systems one has $C'= A+B$ and its experimental indistinguishability from $C$ means again that the states are much more coarse-grained than the observables.

However, the test B has important practical advantages in comparison with the test A. Namely, it is based on two measurement's settings instead  of three and employs inequalities instead of equalities and hence does not require a fine tuning.

\section{Relation to von Neumann theorem}
The main problem with the (in)famous no-go von Neumann theorem is that one cannot find it. The statement from the von Neumann book \cite{vN} which is referred in \cite{MZ} after \cite{Bell, Me} as von Neumann theorem is in fact a combination of the real theorem about the nonexistence of dispersion-free states in quantum mechanics with rather loose remarks on the nonexistence of hidden variable models  (HVM) reproducing quantum mechanical predictions. The mentioned theorem is proved using rather explicit properties of the Hilbert space formalism  instead of general axioms discussed also in the von Neumann book \cite{vNth}.  Unfortunately, von Neumann did not define precisely what was his meaning of HVM what prohibits a simple transformation of his remarks into a theorem.\\  
To continue discussion I can formulate the following hypothesis based on my understanding of \cite{MZ,Bell, Me, vN}:\\
 \emph{Theorem A is the best approximation to what is called in the literature von Neumann no-go theorem}.\\
Under this hypothesis I can agree that the confidence levels of the test
based on the "von Neumann theorem" and the test based on the Theorem B (used in \cite{AV}) are similar (compare statement(I)), although important practical advantages of the later should be acknowledged.

\section{Counterexamples to no-go theorems for HVM}

It is instructive to discuss briefly the structure of some models which might be considered as "counterexamples to no-go theorems for HVM". They should provide "embeddings" of quantum theory into certain classical ones and were mentioned in \cite{MZ}. \\

\emph{Bell's HVM for a qubit}\\
Any pure qubit's state $\rho=\rho^2 = 1/2(I + {\vec k}\cdot{\vec \sigma}), |{\vec k}|=1$, and any qubit's observable $A = a_0I + {\vec a}\cdot{\vec \sigma}$ can be represented by the following probability distribution $p$ and the function $F$ on the 
phase space $\{{\vec m},{\vec n}\}$ which is a Cartesian product of two unit spheres:
\begin{equation}
\rho \equiv p({\vec m},{\vec n}) = \delta ({\vec n}-{\vec k}), \quad  A\equiv F({\vec m},{\vec n})= \bigl\{ a_0 + |{\vec a}|\quad \mathrm{if}\quad({\vec m}+{\vec n})\cdot{\vec a}> 0;\quad a_0 - |{\vec a}|\quad \mathrm{otherwise}\bigr\} ,
\label{spin}
\end{equation}
where $ a_0 \pm |{\vec a}|$ are eigenvalues of $A$. Indeed, one can check that
\begin{equation}
\int d{\vec m}d{\vec n}\, p({\vec m},{\vec n})F({\vec m},{\vec n})= a_0 + {\vec k}\cdot{\vec a}= \mathrm{Tr}(\rho A) .
\label{spin1}
\end{equation}
This model is not minimal and the discussed in the Section 2 symmetry between states and observables is strongly violated. The allowed probability distributions are perfectly localized in $\vec{n}$ and uniform with respect to ${\vec m}$ while the allowed observables are equally sensitive to  both variables.

\emph{HVM of everything }\\
As the set of hidden variables one takes a Cartesian product of the outcomes sets for all relevant observables, disregarding any algebraic relations between them, and as the probability the product of individual probability measures computed from any theory \cite{MZ}. This "model" has predictive power equal to zero and can be treated only as a meaningless  "interpretation" .

\emph{Phase space model}\\
Using an overcomplete set of coherent vectors $\{|\alpha\rangle\}$ one can represent any density matrix $\rho$ by a probability distribution on the phase space (Q-representation)
and any observable $A$ by a phase space function (P-representation) \cite{PQ}
\begin{equation}
\rho \equiv p(\alpha) =\langle\alpha|\rho|\alpha\rangle, \quad  A\equiv F(\alpha)\quad\mathrm{such\ that}\quad A= \int d^2\alpha\,F(\alpha)|\alpha\rangle\langle\alpha|.
\label{pq}
\end{equation}
Although, the classical-like formula holds
\begin{equation}
\mathrm{Tr}(\rho A)= \int d^2\alpha\,p(\alpha)F(\alpha),
\label{pq1}
\end{equation}
this representation is not a HVM  because the values of the function $F(\alpha)$ do not coincide
with the eigenvalues of $A$ corresponding to measurement outcomes. One should notice again the asymmetry between states and observables in this representation. Probability distributions $p(\alpha)$ are "fuzzy" (Heisenberg relations) while $F(\alpha)$ could be even a distribution more singular then Dirac delta.

All those "counterexamples" are not minimal classical models, in the sense of Section 2, moreover they possesses other serious flaws in their mathematical and logical structure.
 
\section{Different faces of Bell's inequality}

Bell's theorem is  a collective name for the vast family of results, some of them meeting the standards of mathematical theorem another ones can be treated as philosophical statements only. The main ingredient is always one of the many forms of Bell's inequality. In the following I use always the so-called Bell-Clauser-Horne-Shimony-Holt inequality. One considers four observables $A_i , B_i, i =1,2$ with the outcomes $1$ or $-1$ under the assumption
that any pair $A_i, B_j$ is simultaneously measurable i.e. the correlation observables  $A_iB_j$ make sense  and yield outputs which are product of the outputs of $A_i$ and $B_j$. Repeating the measurements for different pairs and a single initial state $\rho$ one can compute the following function of  state 
\begin{equation}
 F(\rho)=\rho (A_1B_1) +\rho (A_1B_2) + \rho (A_2B_1)-\rho (A_2B_2)\ .
\label{bell}
\end{equation}
The famous BCHSH inequality reads
\begin{equation}
 |F(\rho)|\leq 2\ .
\label{bell1}
\end{equation}
I discuss now three examples of assumptions which lead to (\ref{bell1}):

\emph{1) Bell's inequality for quantum separable states}\\
Consider a quantum model of bipartite system  with $\{A_i\}$ and $\{B_i\}$ being observables of two different subsystems. Then for all separable states,
i.e. states of the form $\rho=\sum _{\alpha} p_{\alpha}\rho_{\alpha}^A\otimes\rho_{\alpha}^B$, the inequality (\ref{bell1}) holds.\\

\emph{2) Bell's inequality for classical probabilistic models}\\ 
For classical systems (defined as in the Section 2) the inequality (\ref{bell1}) holds for all observables with outcomes $1$ or $-1$ \cite{Lan,Str,APV}.\\

\emph{ 3)Bell's inequality for local realistic models}\\ 
Here the notions of \emph{locality} and \emph{realism} may have different mathematical representations \cite{mathbell} and sometimes the assumption of \emph{free will} is added \cite{stanford, MZ1}.
\par
The case 1) is the least controversial and the most useful in quantum information.
The second one suggests a quantumness test in the spirit
of the examples in Section 2 and for the class of problems Q1. At the first sight, it seems that such a test is  much stronger the the test B as it is enough to show violation of the inequality (\ref{bell1}) for a single state only. However, one should remember that even in the classical theory the observables are jointly measurable only as  abstract idealized objects. In real experiments, concerning for example solid state or atomic implementations of qubits, one should check whether concrete measuring devices acting simultaneously do not interfere with each other introducing unwanted correlations. To exclude this, one needs additional tests with different initial states and observable settings.
Therefore, for practical applications such quantumness tests based on Bell's inequalities are much more involved than single particle tests.

Finally, one can assume that the case 3) is what is really meant by "...\emph{any} classical probabilistic model..." in \cite{MZ}.
As stated , for example in \cite{MZ1} \emph{locality}  means that "...events and actions in Alice's lab cannot influence directly simultaneous events in Bob's lab and his acts...". This is not the case for the most interesting experimental situations concerning controversial  macroscopic quantum systems (superconducting qubits, coupled BEC's , Rydberg atoms,) where Alice and Bob must share the same lab (see discussion of the previous case). To deal with these important and urgent questions one should first apply more feasible single system tests like those proposed in \cite{AV, APV}(compare statement II).

\section{ Example}
In \cite{B} the experiment realizing partially \cite{Ex} test B for a single photon is described. The author of \cite{MZ} proposed to use instead of a single photon source a macroscopic classical beam of light to show that for a macroscopic system the  "quantumness effect"  
can be also observed. Unfortunately, the presented conclusions of his analysis are not correct. A beam is a physical system described by the formalism of second quantisation. For any single-photon observable $A = \sum_j \alpha_j |\phi_j\rangle\langle\phi_j|$ there exists a second quantization observable $\Gamma(A)$ acting on the Fock space and given by
\begin{equation}
\Gamma(A) = \sum_j \alpha_j a^{\dagger}(\phi_j) a(\phi_j)
\label{2q}
\end{equation}
where $a(\phi_j) ,( a^{\dagger}(\phi_j))$ is an annihilation (creation) operator corresponding to a mode of radiation (equivalently, a single photon wave function normalized to 1) $\phi_j$. The outcomes of the observables (\ref{2q}) are given by $ \alpha_j n_j $ , $n_j = 0,1,2,...$ . On should remember that only the \emph{additive observables} (\ref{2q}) and their functions (e.q. moments) can be measured in linear optics experiments.\\
The second quantization map $A\to \Gamma(A)$ preserves the order i.e.:
\begin{equation}
0\leq A\leq B ,\quad A^2 \nleq B^2 \Rightarrow 0\leq \Gamma(A)\leq \Gamma(B),\quad  \Gamma(A^2) \nleq \Gamma(B^2) .
\label{qine}
\end{equation}
but not the algebraic relations, e.g. $\Gamma(A^2)\neq \bigl(\Gamma(A)\bigr)^2$.\\
The quantumnes test B involves squares of the observables like $\bigl(\Gamma(A)\bigr)^2$ with the outcomes $(\alpha_j n_j)^2$ \emph{but not} $\Gamma(A^2)$ with the outcomes $\alpha^2_jn_j$. Notice that $\bigl(\Gamma(A)\bigr)^2$ coincides with $\Gamma(A^2)$ only on the vacuum and 1-photon sector of the Fock space. Using the formula 
\begin{equation}
\bigl(\Gamma(A)\bigr)^2 = \sum_{i,j} \alpha_i \alpha_j a^{\dagger}(\phi_i) a(\phi_i) a^{\dagger}(\phi_j) a(\phi_j)=\sum_{i,j} \alpha_i \alpha_j a^{\dagger}(\phi_i) a^{\dagger}(\phi_j)a(\phi_i)) a(\phi_j)+ \Gamma (A^2).
\label{2Q}
\end{equation}
one can compute the mean value of the relevant observables in a coherent state $\Phi(\xi)$. Here $\xi$ is interpreted as a single-photon wave function normalized to the averaged number of photons , $\langle \xi|\xi\rangle = N$. For large $N$ the coherent state describes a quantum state of a macroscopic light beam
determined by $\xi$, interpreted now as the classical electromagnetic field.  One obtains the formulas
\begin{equation}
\langle \Phi(\xi),\Gamma(A)\Phi(\xi)\rangle =\langle \xi,A \xi\rangle 
\label{ine}
\end{equation}
and
\begin{equation}
\langle \Phi(\xi),\bigl(\Gamma(A)\bigr)^2\Phi(\xi)\rangle =(\langle \xi,A \xi\rangle)^2 + \langle \xi,A^2 \xi\rangle\ .
\label{ave}
\end{equation}

The first term on the RHS of (\ref{ave}) is of the order of $N^2$ while the second is of the order of $N$. Therefore, only for a weak beam, i.e. $N <<1$
the second term dominates and with the choice of single-photon observables $0\leq A\leq B $ and  $A^2 \nleq B^2$ yields the "violation of classicality". For macroscopic beams $(N>>1)$ the first term dominates and the classical order relations are preserved (compare with the discussion of "macroscopic entanglement" in \cite{RA} or "additive observables" in \cite{APV}). Therefore, only experiments with single photons can show \emph{directly} deviations from classical probabilistic model. Of course, in such experiments any single photon  source is fine \cite{XIX} and the results has nothing to do with the "statistical properties of the parametric down conversion"  (compare statement III) applied in the experimental setting of \cite{B}.
\par
The interesting aspect of linear optics experiments with light beams is that completely deterministic macroscopic experiments on the macroscopic objects can
provide \emph{indirectly} information about the quantum nature of the \emph{underlying microscopic constituents}. It is possible under the hypothesis which can be called  \emph{Newton's model of light}:\\
\emph{A light beam consists of  noninteracting indistinguishable particles which interact independently with the measuring apparatus}.\\
The unique nature of photons, in particular the combination of bosonic statistics,  mass and charge equal to zero and strong interaction with matter allows to prepare
states satisfying Newton's hypothesis and yielding high values of the outcomes $\alpha_j n_j$ which produce macroscopic effects \cite{cal}. Hence, 
we can interpret a measurement performed on a single system - a light beam - as equivalent to a sequence of many independent measurements performed on a single photon prepared always in a fixed state .
\par
As a consequence sir George Gabriel Stokes, who showed in 1852 that a state of light beam polarization is described by only four parameters, should be recognized as the discoverer of quantum mechanics \cite{stokes}. Indeed, under the Newton's hypothesis the Stokes result implies that the polarization's state of a single photon must be described
by three parameters represented by a vector in the interior of the qubit's Bloch sphere (called Poincar\'e sphere in polarization optics). If polarization would be a classical system its states should be represented by an infinitely dimensional simplex of probability measures instead of the 3-dimensional ball.

\emph{Acknowledgements.} The author thanks Marco Piani, Micha\l Horodecki and Nick Van Ryn  for
discussions.   The work is  supported by the Polish research network LFPPI.


\begin{thebibliography}{99}
\bibitem{phil}
The relations between classical probabilistic models, hidden variable models,  and local realistic models seem to be quite subtle and are usually discussed in philosophical and not mathematical language using the words like "reality", "contextual", "non-contextual" and even "contingency"
\cite{stanford}. An often ignored difference between "model" and "interpretation" is another source of confusions.
\bibitem{alg}
The reader unfamiliar with the algebraic approach can always consider  two models: the classical one with the algebra of functions and states as probability measures and the quantum one with the algebra of all bounded operators on the Hilbert space and states being density operators (matrices).
\bibitem{obs}
To avoid a complicated notation I use the same symbol and the same name for an "observable" ("state") as a physical, operational notion and an "observable" ("state")
as its mathematical representation.
\bibitem{dual}
The symmetry between states and observables appears in the mathematical framework in the form of duality between states and observables.
The states can be treated as linear functionals of observables and vice versa. The alternative- $C^*$ versus von Neumann algebras -
reflects this feature on a very abstract level.
\bibitem{Ex}
The missing part of the quantumness test in the experimental setting of \cite{B} is the verification
of the inequality $\rho(A) \leq \rho(B)$ for a large set of initial states.
\bibitem{XIX}
In fact, this type of experiment could be performed in XIX century if only a well-trained human eye could detect a single photon.
\bibitem{cal}
One does not need to count photons, calorimetry is sufficient.
\bibitem{stokes}
In the same sense as Christopher Columbus was the discoverer of America.
\end{thebibliography}
\end{document}